\documentclass[accepted=2022-11-07,aps,10pt,longbibliography,noeprint,a4paper]{quantumarticle}
\pdfoutput=1

\usepackage{amsmath,amsfonts,amssymb,amsthm,bbm,graphicx,enumerate,times}
\usepackage{mathtools}
\usepackage[usenames,dvipsnames]{color}
\usepackage{todonotes} %[disable]
\usepackage{comment}
\usepackage{float}
\usepackage{hyperref}
\usepackage{tikz}
\usepackage{graphicx}
\usepackage{bm}
\usetikzlibrary{shapes,positioning}
\usepackage{etoolbox}

 \newtheorem{theorem}{Theorem}
 
 \newtheorem{conjecture}{Conjecture}

 \newtheorem{lemma}[theorem]{Lemma}
 \newtheorem{corollary}[theorem]{Corollary}
 \newtheorem{definition}[theorem]{Definition}

\definecolor{henrik}{rgb}{.8,.3,0}

\newcommand{\mc}[1]{\mathcal{#1}}

\newcommand{\mb}[1]{\mathbb{#1}}

\newcommand{\tr}{\mathrm{Tr}} %old
\newcommand{\Tr}{\mathrm{Tr}} %new

\newcommand{\id}{\mathbbm{1}}

\newcommand{\N}{\mb{N}}

\newcommand{\norm}[1]{\left\Vert #1 \right\Vert}

\newcommand{\ket}[1]{\left.\left|{#1}\right.\right\rangle}

\newcommand{\bra}[1]{\left.\left\langle{#1}\right.\right|}

\newcommand{\ketbra}[2]{\ket{#1} \!\! \bra{#2}}

  \newcommand{\proj}[1]{\ketbra{#1}{#1}}
\renewcommand{\vec}[1]{\pmb{#1}}

\begin{document}
\title{Correlations in typicality and an affirmative solution to the exact catalytic entropy conjecture}
 
\author{Henrik Wilming}
%%\email{henrikw@phys.ethz.ch}
\affiliation{Leibniz Universit\"at Hannover, Appelstra\ss e 2, 30167 Hannover, Germany}
%\affiliation{Institute  for  Theoretical  Physics,  ETH  Zurich,  8093  Zurich,  Switzerland}
%\author{P.\ Boes}
%\affiliation{\fu}
\begin{abstract}
	It is well known that if a (finite-dimensional) density matrix $\rho$ has smaller entropy than $\rho'$, then the tensor product of sufficiently many copies of $\rho$ majorizes
	a quantum state arbitrarily close to the tensor product of correspondingly many copies of $\rho'$.
	In this short note I show that if additionally $\mathrm{rank}(\rho)\leq \mathrm{rank}(\rho')$, then $n$ copies of $\rho$ also majorize a state where all single-body marginals are \emph{exactly identical} to $\rho'$
	but arbitrary correlations are allowed (for some sufficiently large $n$). 
	An immediate application of this is an affirmative solution of the exact catalytic entropy conjecture introduced by Boes et al. [PRL 122, 210402 (2019)]:
	If $H(\rho)<H(\rho')$ and $\mathrm{rank}(\rho)\leq \mathrm{rank}(\rho')$ there exists a finite dimensional density matrix $\sigma$ and a unitary $U$ such that $U\rho\otimes \sigma U^\dagger$ has marginals $\rho'$ and $\sigma$ \emph{exactly}.
	All the results transfer to the classical setting of probability distributions over finite alphabets with unitaries replaced by permutations.
\end{abstract}
\maketitle
\section{Introduction and results}
\begin{figure}[t!]
		\includegraphics[width=8cm]{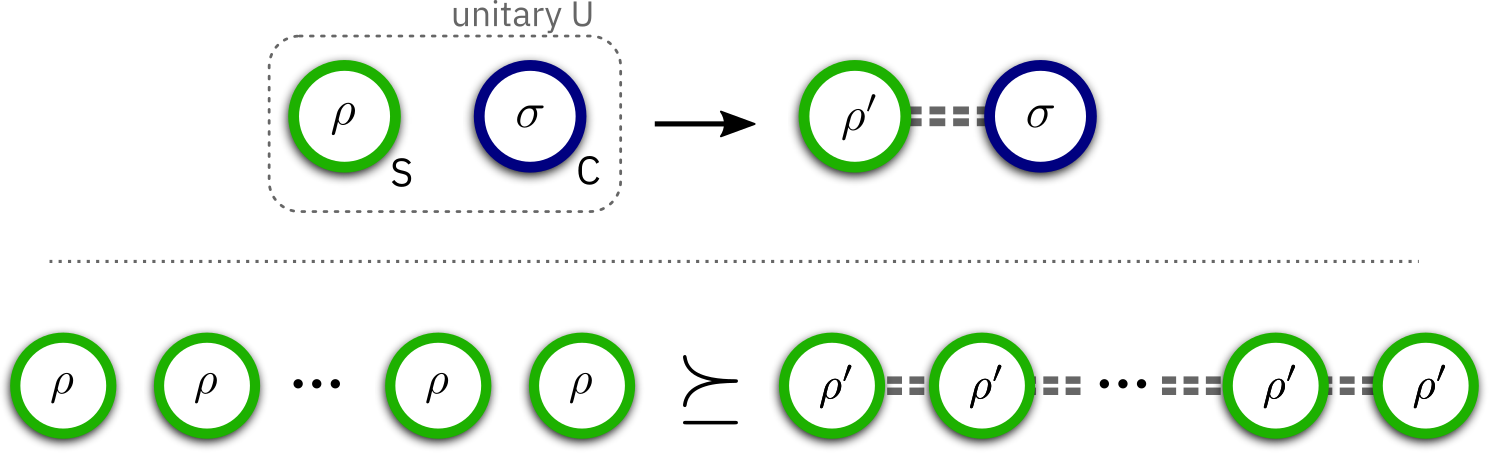}
	\caption{Top (CEC): A unitary operation $U$ is applied to systems $S$ and $C$ in the state $\rho\otimes\sigma$. The resulting reduced state on $S$ is exactly $\rho'$, while the reduced state on $C$ is preserved exactly, but correlated to $S$ (indicated by the dashed lines). Both $\sigma$ and $U$ may depend on $\rho$ and $\rho'$.
		Bottom (Lemma~\ref{lemma:main}): For some finite $n$, the state $\rho^{\otimes n}$ majorizes a state $\rho'_n$ with all marginals equal to $\rho'$.
		The results of this paper show that if $\rho$ and $\rho'$ are not unitarily equivalent, then both situations occur if and only if $H(\rho)<H(\rho')$ and $\mathrm{rank}(\rho)\leq \mathrm{rank}(\rho')$. Both results transfer to the classical setting as well.}
		\label{fig:main}
\end{figure}

The exact catalytic entropy conjecture (CEC) is the following conjecture from Ref.~\cite{Boes2018b} about the von~Neumann entropy $H(\rho):=-\tr[\rho\log(\rho)]$ ($\log$ with base $2$) illustrated in Fig.~\ref{fig:main} (top):
\begin{conjecture}[CEC]
	Let $\rho,\rho'$ be $d$-dimensional density matrices that are not unitarily equivalent. 
	Then there exists a finite-dimensional density matrix $\sigma$ and a unitary $U$ such that
	\begin{align}
		\tr_1[U\rho\otimes\sigma U^\dagger] = \sigma,\tr_2[U\rho\otimes\sigma U^\dagger]=\rho'
	\end{align}
	if and only if $H(\rho)<H(\rho')$ and $\mathrm{rank}(\rho)\leq \mathrm{rank}(\rho')$.
\end{conjecture}
An affirmative solution to the CEC provides a clean single-shot interpretation of von~Neumann entropy without external randomness using the "catalyst" $\sigma$,
see Refs.~\cite{Boes2018b,Wilming2021} for discussions of the conjecture and their physical implications.
In Ref.~\cite{Wilming2021}, an approximate version of the CEC was proven. 
It follows directly from the proof, which itself is similar to earlier constructions in Refs.~\cite{Duan2005a,Feng2006,Shiraishi2020}, that the CEC is true if the following Lemma holds true, illustrated in Fig.~\ref{fig:main} (bottom).
For completeness, we provide the explicit construction of $U$ and $\sigma$ in the CEC based on the following Lemma in the appendix.
\begin{lemma}\label{lemma:main}
	Let $\rho,\rho'$ be $d$-dimensional density matrices that are not unitarily equivalent.
	Then there exist $n\in \N$ and a density matrix $\rho'_n$ on $n$ copies of $\mathbb C^d$ such that
	\begin{align}
		\rho^{\otimes n} \succeq \rho'_n,\quad \tr_{\{1,\ldots,n\}\setminus i }[\rho'_n]=\rho'\ \forall i=1,\ldots,n
	\end{align}
	if and only if $H(\rho)< H(\rho')$ and $\mathrm{rank}(\rho)\leq \mathrm{rank}(\rho')$.
\end{lemma}
Here, for any two density matrices $\rho$ and $\sigma$, $\rho \succeq \sigma$ ("$\rho$ majorizes $\sigma$") means 
that there exists a finite collection of unitaries $U_i$ with associated probabilities $p_i$ such that $\sigma = \sum_i p_i U_i \rho U_i^\dagger$. 
More explicitly, there exists a unitary $V$ such that $\sigma = \mc D_\sigma[V\rho V^\dagger]$, where $\mc D_\sigma$ denotes the \emph{decoherence-chanel} in the eigenbasis $\{\ket j\}$ of $\sigma=\sum_j q_j \proj{j}$. 
It acts on any operator $X$ as $\mc D_\sigma[X] = \sum_j \proj{j}X\proj{j}$, but can be represented as $\mc D_\sigma[X] = \frac{1}{d}\sum_i U_iXU_i^\dagger$ for a suitable set of $d$ unitaries $U_i$. 
 
An equivalent definition of majorization in terms eigenvalues is as follows (by the Schur-Horn Lemma and Birkhoff's Theorem, see for example Refs.~\cite{Bhatia1997,Marshall2011}): if $\vec \lambda_\rho$ denotes the vector of eigenvalues of $\rho$ (including multiplicities) and $\vec \lambda_\sigma$ the corresponding vector of eigenvalues of $\sigma$, then there exists a finite collection of permutation matrices $\Pi_i$ and a probability distribution $q_i$ such that $\vec \lambda_\sigma = \sum_i q_i \Pi_i \vec \lambda_\rho$. 
This latter characterization of majorization is also the appropriate one for general probability vectors.

%%%%%%%%%%%%%%%%%%%%%%
%%%%%%%%%%%%%%%%%%%%%5
In words Lemma~\ref{lemma:main} reads: if the entropy of $\rho'$ is higher than that of $\rho$ (and the rank not smaller), then for some finite $n$ the state $\rho^{\otimes n}$ majorizes a state whose one-body marginals are all equal to $\rho'$. 
It is well known that for sufficiently large $n$ the state $\rho^{\otimes n}$ majorizes an $\epsilon$-approximation of ${\rho'}^{\otimes n}$. Lemma~\ref{lemma:main} instead shows that $\rho^{\otimes n}$ majorizes a state whose marginals coincide exactly with $\rho'$, but where arbitrary correlations are allowed.  

Our aim here is to prove Lemma~\ref{lemma:main} and thereby the exact catalytic entropy conjecture. In fact, the arguments below can also all be transferred to the classical setting, where density matrices are replaced by finite-dimensional probability vectors and unitary transformations by permutations of the entries of the vectors.
Using the corresponding classical construction in Ref.~\cite{Wilming2021} this implies that the classical version of the CEC also holds (contrarily to some prior beliefs, see Supplemental Material of Ref.~\cite{Boes2018b}):
\begin{theorem}[classical CEC]\label{thm:classical}
	Let $\vec p,\vec p' \in \mathbb R^d$ be probability vectors that are not related by a permutation of their entries.  Then there exists a finite-dimensional probability vector $\vec q$ and a permutation matrix $\Pi$ such that
	\begin{align}
		\tr_{1}[\Pi \vec p\otimes \vec q] = \vec q,\quad \tr_2[\Pi \vec p\otimes \vec q] = \vec p'
	\end{align}
	if and only if $H(\vec p)<H(\vec p')$ and $\mathrm{rank}(\vec p)\leq \mathrm{rank}(\vec p')$, with the Shannon entropy $H(\vec p) = -\sum_i p_i \log(p_i)$.
\end{theorem}
In the formulation of Theorem~\ref{thm:classical}, $\tr_{x}$ denotes marginalization over tensor-factor $x$ and $\mathrm{rank}(\vec x)$ denotes the number of non-zero elements of a vector $\vec x$. 

In fact, contrary to the usual situation where results in quantum theory imply corresponding classical results, Theorem~\ref{thm:classical} implies the quantum version of the CEC but not vice-versa. However, as in the approximate case, the dimension required for the catalyst can be significantly smaller in the quantum case compared to the classical case, see Ref.~\cite{Wilming2021}.

\section{Resource theories}
In Refs.~\cite{Mueller2017,Shiraishi2020,Kondra2021,LipkaBartosik2021} catalysts that become correlated to the system of interest have been used to provide single-shot interpretations of the non-equilibrium free energy in quantum thermodynamics and the entanglement entropy in the framework of local operations and classical communication. 
These results were approximate, in the sense that the output state of a protocol only approximated the desired state up to arbitrary accuracy as the catalyst dimension diverged. 
Our result may suggest that a \emph{strict} decrease in the free energy or entanglement entropy could imply the possibility of an exact transformation with a finite-dimensional catalyst also in those settings. 
However, Ref.~\cite{Rubboli2021} showed already that this is not the case, at least if one demands that the final correlations between catalyst and system can be made arbitrarily small. 
In the setting of this note, we cannot make the latter assumption since the mutual information between catalyst and final state of the system is always given by $H(\rho')-H(\rho)$.
It would be interesting to know whether in the context of quantum thermodynamics a strict decrease in free energy allows for an exact conversion with finite-dimensional catalyst if one does not put a condition on the final correlations. 
Indeed, Ref.~\cite{Rethinasamy2020} showed that this is true in the quasi-classical case with the additional assumption that the thermal state only has rational entries in its eigenbasis. 

\section{Open problems}
Lemma~\ref{lemma:typ} does not claim that the state $\rho'_n$ is close to ${\rho'}^{\otimes n}$. 
It is an open problem whether one can both arrange that the marginals of $\rho'_n$ are exactly equal to $\rho'$ \emph{and} $\rho'_n$ is arbitrarily close to ${\rho'}^{\otimes n}$ in trace-distance while keeping the majorization condition $\rho^{\otimes n}\succeq \rho'_n$. If true, this statement could likely have many further applications in physics and information theory.  

A second open problem concerns the case $H(\rho)=H(\rho')$. It is known from lower bounds on the catalyst dimension that in this case a finite-dimensional catalyst in general cannot be used to implement the state-transition exactly \cite{Boes2022,Blakaj2021}. On the other hand, continuity of von Neumann entropy implies that it can be implemented arbitrarily well as the catalyst dimension diverges. However, it is presently unknown whether a catalyst with properly infinite dimensional density matrix could be used to implement the state-transition $\rho\rightarrow \rho'$ exactly in this case. 

\section{The proof}
We formulate the proof in the quantum language, but, as mentioned above, everything transfers to the classical case by replacing density matrices with probability vectors. 
\begin{definition}
	For every finite-dimensional density matrix $\rho$ and $n\in\N$, we define the following set:
	\begin{align}
		S_n(\rho) := \{ \frac{1}{n}\sum_i \sigma_{\{i\}} \mid \rho^{\otimes n} \succeq \sigma, \sigma\geq 0,\Tr[\sigma]=1  \},
	\end{align}
	where $\sigma_{\{i\}}=\tr_{\{1,\ldots,n\}\setminus i }[\sigma]$. 
\end{definition}

Since randomly permuting subsystems is a random unitary process (in the sense of majorization), whenever $\rho'\in S_n(\rho)$ there exists a state $\rho'_n$ with $\rho^{\otimes n}\succeq \rho'_n$ as in Lemma~\ref{lemma:main}. 
So we want to show that whenever $H(\rho')>H(\rho)$ and $\mathrm{rank}(\rho)\leq \mathrm{rank}(\rho')$ then $\rho'\in S_n(\rho)$ for some $n\in N$.
\begin{lemma}
	For every $\rho$ on $\mathbb C^d$, $S_n(\rho)$ is a compact convex set within the set of density matrices on $\mathbb C^d$ and $\id/d \in S_n(\rho)$. 
	Furthermore, $S_n(\rho)$ has full dimension within the set of density matrices.  
\end{lemma}
	\begin{proof}
		The set of density matrices $\sigma$ such that $\rho^{\otimes n}\succeq \sigma$ is closed and convex and hence compact. $S_n(\rho)$ is convex and compact, because it is its image under a linear (and hence continuous) map. 
		 It has full dimension since in particular it includes an open ball around $\id/d$ (and hence also $\id/d$).
	\end{proof}
We now introduce a form of typicality. Note that the statement is not in the usual form, as we claim a statement for all density matrices with certain entropies.
In other words, we require an error bound that is uniform over \emph{all} density matrices with a certain minimal entropy.
To state the Lemma, we define the distance $d(\rho,S)$ between a density matrix $\rho$ and a set of density matrices $S$ as $d(\rho,S) = \inf_{\sigma \in S}d(\rho,\sigma)$, where $d(\rho,\sigma) = \frac{1}{2}\norm{\rho-\sigma}_1$ denotes the trace-distance.

\begin{lemma}[Typicality]\label{lemma:typ} For every $\delta>0$ and $\epsilon>0$ there exists $n_\epsilon \in N$ such that for all states $\rho'$ with $H(\rho')>H(\rho)+2\delta$, 
	\begin{align}
		d(\rho',S_{n_\epsilon}(\rho)) \leq \epsilon.
	\end{align}
	\begin{proof}
		In general, a density matrix $\rho$ majorizes an $\epsilon$-approximation of $\rho'$ if $H^{\epsilon/2}_{\mathrm{max}}(\rho) < H^{\epsilon/2}_{\mathrm{min}}(\rho')$, where $H^{\epsilon}_{\mathrm{max}}$ is the smooth max-entropy and $H^{\epsilon}_{\mathrm{min}}$ the smooth min-entropy, see \cite{Renner2005,Tomamichel2016}.
		By Theorem 1 of Ref.~\cite{Holenstein2011} (in Ref.~\cite{Holenstein2011}, the bounds are stated for conditional entropies of classical random variables, but in the unconditional case, they generalize immediately to the quantum case), it follows that for every $\delta\geq 0$ we have
		\begin{align}
			H^{\epsilon/2}_{\mathrm{max}}({\rho}^{\otimes n}) &\leq n( H(\rho) + \delta),\\
			H^{\epsilon/2}_{\mathrm{min}}({\rho'}^{\otimes n}) &\geq n(H(\rho') - \delta)
		\end{align}
		for $\epsilon =2^{-\frac{n\delta^2}{2 \log^2(d+3)}}$ (which does not depend on $\rho'$ or $\rho$) \footnote{We define the smoothing of the entropies in terms of trace-distance instead of $1$-norm, which accounts for a factor of $1/2$.}.
	Thus, if we take
		\begin{align}
			n_\epsilon = \left\lceil \frac{2\log^2(d+3)}{\delta^2}\log(\frac{1}{\epsilon}) \right\rceil,
		\end{align}
	we find that $\rho^{\otimes n_\epsilon}$ majorizes an $\epsilon$-approximation $\rho'_{\epsilon,n_\epsilon}$ of ${\rho'}^{\otimes n_\epsilon}$ as long as $H(\rho')>H(\rho)+2\delta$.
	By the triangle inequality and monotonicity of the trace-distance under quantum channels, we find:
		\begin{align}
			d(\frac{1}{n_\epsilon} \sum_{i=1}^{n_\epsilon} \tr_{\{1,\ldots,n_\epsilon\}\setminus i}[\rho'_{\epsilon,n_\epsilon}],\rho') \leq \epsilon.
		\end{align}
		Thus $d(\rho',S_{n_\epsilon}(\rho))\leq \epsilon$.
	
	\end{proof}
\end{lemma}
The crucial additional ingredient to our proof apart from the typicality Lemma is the following simple observation about convex geometry.

\begin{lemma}\label{lemma:convex}
	Let $y\in \mathbb R^d$. Let $B_y(\delta)$ denote an open ball of radius $\delta$ around $y$ and let $C_\epsilon$ be a parametrized family of closed and convex sets with the following property:
	\begin{align}	
		\forall x\in B_y(\delta)\quad d(x,C_\epsilon) \leq \epsilon,
	\end{align}
	where $d$ is the euclidean distance on $\mathbb R^d$. 
	Then $y\in C_\epsilon$ for all $\epsilon<\delta/2$.
	\begin{proof}
		Let $\epsilon<\delta/2$ and suppose contrarily that $y\notin C_\epsilon$. 
		Let $c$ be the element in $C_\epsilon$ closest to $y$ (which is unique because of convexity of $C_\epsilon$).
		But since $y$ is the center of the ball, there must then be an element $z\in B_y(\delta)$ with $\epsilon \geq d(z,C_\epsilon) \geq d(z,y) > \delta/2$, which is a contradiction. 		To find $z$, shoot a ray through $y$ and $c$. The part of the ray inside of $B_y(\delta)$ has length at least $\delta$, so in particular there must be a corresponding element $z$ with $d(z,c)>\delta/2$.
	\end{proof}
\end{lemma}
\begin{corollary}\label{cor:convex}
Let $C$ be a full-dimensional convex set in a normed, finite-dimensional, real vector space $V$ and let $C_\epsilon\subseteq V$ be a parametrized family of closed and convex sets with the following property:
	\begin{align}
		\forall x \in C\quad d(x,C_\epsilon)<\epsilon,
	\end{align}
	where $d$ is the distance induced by the norm.
	Then for any  $y\in\mathrm{int}\,C$ there exists an $\epsilon_0$ such that $y \in C_\epsilon$ for all $\epsilon<\epsilon_0$.
	\begin{proof}
		Since $y$ is an interior point, there is an open ball $B_y(\delta)$ around it and we can apply Lemma~\ref{lemma:convex} (after embedding everything in some $\mathbb R^d$).
	\end{proof}
\end{corollary}

We can now prove Lemma~\ref{lemma:main}: 
We start with the "if" part.
By unitary freedom and because $\mathrm{rank}(\rho)\leq \mathrm{rank}(\rho')$, we can assume that $\mathrm{supp}(\rho)\subseteq \mathrm{supp}(\rho')$. 
Indeed, the rank-condition simply means that the dimension of the support of $\rho$ is not larger than the dimension of the support of $\rho'$. 
Furthermore, we can shrink Hilbert-space to the support of $\rho'$, so that $\rho'$ has full rank ($\rho'>0$). We will do so in the following. 
So let $H(\rho')-H(\rho) = 2\delta>0$ and $\rho'>0$. Then $\rho'$ is an interior point of the convex set of all density matrices since it has full rank. 
Moreover, it has entropy $H(\rho')>H(\rho)+\delta$ and von Neumann entropy is continuous. 
Therefore, there is an open ball of states around it (within the set of density matrices) all of which have entropy $> H(\rho)+\delta$.
Hence it is an interior point of the convex set $C$ of density matrices with entropy $\geq H(\rho)+\delta$ and $C$ is full-dimensional.
For any $\epsilon>0$ set $C_\epsilon = S_{n_\epsilon}(\rho)$ with $n_\epsilon$ from Lemma~\ref{lemma:typ}. Then by the same Lemma, the conditions of Corollary~\ref{cor:convex} are fulfilled if we set $y=\rho'$. Hence there exists some $n$ such that $\rho'\in S_n(\rho)$. 

For the "only if" part let us start with the condition on the von~Neumann entropy: We use that $H$ is sub-additive, $H(\rho_{12}) \leq H(\rho_1) + H(\rho_2)$ with equality if and only if $\rho_{12} = \rho_1\otimes\rho_2$, and strictly Schur-concave, meaning that if $\rho\succeq \rho'$ but $\rho$ and $\rho'$ are not unitarily equivalent, then $H(\rho)<H(\rho')$. So suppose that $\rho^{\otimes n} \succeq \rho'_n$ and $\rho$ and $\rho'$ are not unitarily equivalent. Then we have
\begin{align}\label{eq:mutual_info}
	n H(\rho) = H(\rho^{\otimes n}) \leq H(\rho'_n) \leq n H(\rho'). 
\end{align}
We now assume $H(\rho)=H(\rho')$ to arrive at a contradiction. Then \eqref{eq:mutual_info} implies $H(\rho'_n)=nH(\rho')$, which in turn implies $\rho'_n = {\rho'}^{\otimes n}$. But then $\rho^{\otimes n} \succeq {\rho'}^{\otimes n}$ and $H(\rho) = H(\rho')$, which by strict Schur-concavity implies that $\rho^{\otimes n}$ and ${\rho'}^{\otimes n}$ are unitarily equivalent. But then $\rho$ and $\rho'$ are unitarily equivalent, which yields the contradiction. Hence $H(\rho)<H(\rho')$.    
To conclude $\mathrm{rank}(\rho)\leq \mathrm{rank}(\rho')$, we use that $H_0(\rho) := \log\left(\mathrm{rank}(\rho)\right)$ is sub-additive \cite{Linden2013} and additive over-tensor factors as well as non-decreasing under majorization. Hence
\begin{align}
	H_0(\rho^{\otimes n}) = n H_0(\rho) \leq H_0(\rho'_n) \leq n H_0(\rho'), 
\end{align}
which finishes the proof. 

\paragraph{Acknowledgements.}
I would like to thank Niklas Galke for rekindling my interest in this problem and pointing out a mistake in a previous version of the argument. 
I would also like to thank  Paul Boes, Thomas Cope, Patryk Lipka-Bartosik, Nelly H.Y. Ng and Reinhard F. Werner for discussions as well as Roberto Rubboli and two anonymous referees for useful feedback on a previous version of the paper.
Support by the DFG through SFB 1227 (DQ-mat), Quantum Valley Lower Saxony, and funding by the Deutsche Forschungsgemeinschaft (DFG, German Research Foundation) under Germanys Excellence Strategy EXC-2123 QuantumFrontiers 390837967 is also acknowledged.

\bibliographystyle{quantum}
\bibliography{bibliography}

\begin{thebibliography}{10}

\bibitem{Boes2018b}
Paul Boes, Jens Eisert, Rodrigo Gallego, Markus~P. Müller, and Henrik Wilming.
\newblock ``Von neumann entropy from unitarity''.
\newblock \href{https://dx.doi.org/10.1103/physrevlett.122.210402}{Physical
  Review Letters {\bf 122}, 210402}~(2019).

\bibitem{Wilming2021}
H.~Wilming.
\newblock ``Entropy and reversible catalysis''.
\newblock \href{https://dx.doi.org/10.1103/physrevlett.127.260402}{Physical
  Review Letters {\bf 127}, 260402}~(2021).

\bibitem{Duan2005a}
Runyao Duan, Yuan Feng, Xin Li, and Mingsheng Ying.
\newblock ``Multiple-copy entanglement transformation and entanglement
  catalysis''.
\newblock \href{https://dx.doi.org/10.1103/PhysRevA.71.042319}{Phys. Rev. A
  {\bf 71}, 042319}~(2005).

\bibitem{Feng2006}
Yuan Feng, Runyao Duan, and Mingsheng Ying.
\newblock ``Relation between catalyst-assisted transformation and multiple-copy
  transformation for bipartite pure states''.
\newblock \href{https://dx.doi.org/10.1103/physreva.74.042312}{Physical Review
  A {\bf 74}, 042312}~(2006).

\bibitem{Shiraishi2020}
Naoto Shiraishi and Takahiro Sagawa.
\newblock ``Quantum thermodynamics of correlated-catalytic state conversion at
  small scale''.
\newblock \href{https://dx.doi.org/10.1103/physrevlett.126.150502}{Physical
  Review Letters {\bf 126}, 150502}~(2021).

\bibitem{Bhatia1997}
Rajendra Bhatia.
\newblock ``Matrix analysis''.
\newblock \href{https://dx.doi.org/10.1007/978-1-4612-0653-8}{Springer New
  York}. ~(1997).

\bibitem{Marshall2011}
Albert~W. Marshall, Ingram. Olkin, and Barry~C. Arnold.
\newblock ``{Inequalities : theory of majorization and its applications}''.
\newblock \href{https://dx.doi.org/10.1007/978-0-387-68276-1}{Springer
  Science+Business Media, LLC}. ~(2011).

\bibitem{Mueller2017}
Markus~P. Müller.
\newblock ``Correlating thermal machines and the second law at the nanoscale''.
\newblock \href{https://dx.doi.org/10.1103/physrevx.8.041051}{Physical Review X
  {\bf 8}, 041051}~(2018).

\bibitem{Kondra2021}
Tulja~Varun Kondra, Chandan Datta, and Alexander Streltsov.
\newblock ``Catalytic transformations of pure entangled states''.
\newblock \href{https://dx.doi.org/10.1103/physrevlett.127.150503}{Physical
  Review Letters {\bf 127}, 150503}~(2021).

\bibitem{LipkaBartosik2021}
Patryk Lipka-Bartosik and Paul Skrzypczyk.
\newblock ``Catalytic quantum teleportation''.
\newblock \href{https://dx.doi.org/10.1103/physrevlett.127.080502}{Physical
  Review Letters {\bf 127}, 080502}~(2021).

\bibitem{Rubboli2021}
Roberto Rubboli and Marco Tomamichel.
\newblock ``Fundamental limits on correlated catalytic state transformations''.
\newblock \href{https://dx.doi.org/10.1103/physrevlett.129.120506}{Physical
  Review Letters {\bf 129}, 120506}~(2022).

\bibitem{Rethinasamy2020}
Soorya Rethinasamy and Mark~M. Wilde.
\newblock ``Relative entropy and catalytic relative majorization''.
\newblock \href{https://dx.doi.org/10.1103/physrevresearch.2.033455}{Physical
  Review Research {\bf 2}, 033455}~(2020).

\bibitem{Boes2022}
Paul Boes, Nelly~H.Y. Ng, and Henrik Wilming.
\newblock ``Variance of relative surprisal as single-shot quantifier''.
\newblock \href{https://dx.doi.org/10.1103/prxquantum.3.010325}{{PRX} Quantum
  {\bf 3}, 010325}~(2022).

\bibitem{Blakaj2021}
Vjosa Blakaj and Michael~M. Wolf.
\newblock ``Transcendental properties of entropy-constrained sets''~(2021).
\newblock  \href{http://arxiv.org/abs/2111.10363}{arXiv:2111.10363}.

\bibitem{Renner2005}
R.~{Renner}.
\newblock ``{Security of Quantum Key Distribution}''.
\newblock PhD thesis.
\newblock ETH Zurich.
\newblock ~(2005).

\bibitem{Tomamichel2016}
Marco Tomamichel.
\newblock ``Quantum information processing with finite resources''.
\newblock \href{https://dx.doi.org/10.1007/978-3-319-21891-5}{Springer
  International Publishing}. ~(2016).

\bibitem{Holenstein2011}
T~Holenstein and R~Renner.
\newblock ``On the randomness of independent experiments''.
\newblock \href{https://dx.doi.org/10.1109/tit.2011.2110230}{{IEEE}
  Transactions on Information Theory {\bf 57}, 1865--1871}~(2011).

\bibitem{Linden2013}
Noah Linden, Mil{\'{a}}n Mosonyi, and Andreas Winter.
\newblock ``The structure of r{\'{e}}nyi entropic inequalities''.
\newblock \href{https://dx.doi.org/10.1098/rspa.2012.0737}{Proceedings of the
  Royal Society A: Mathematical, Physical and Engineering Sciences {\bf 469},
  20120737}~(2013).

\end{thebibliography}
\phantom{x}
\appendix
\onecolumngrid
\section{The construction of the catalyst and unitary in the CEC}
In this section, we use the construction of the catalyst and unitary in Ref.~\cite{Wilming2021} to prove the existence of $U$ and $\sigma$ in the CEC using Lemma~\ref{lemma:main}.
We assume that $\rho,\rho',n$ and $\rho'_n$ are given as in Lemma~\ref{lemma:main}.
Let $\rho' = \sum_j p'_j \proj{j}$ be the spectral decomposition of $\rho'$ 
and $\mc D_\rho'[\cdot]$ the decoherence-channel in the eigenbasis of $\rho'$.
We can without loss of generality assume that $\mc D^{\otimes n}_{\rho'}[\rho'_n] = \rho'_n$, 
since if $\rho^{\otimes n} \succeq \rho'_n$, then $\mc D^{\otimes n}_{\rho'}[\rho'_n]$ is also majorized by $\rho^{\otimes n}$ and locally identical to $\rho'$.
Thus there exists a unitary $V$ such that
\begin{align}
	\rho'_n = \mc D_{\rho'}^{\otimes n} [V \rho^{\otimes n} V^\dagger].
\end{align}
For ease of notation, let us write $\chi = V \rho^{\otimes n} V^\dagger$ and $\chi_{[1:l]}$ for the reduced state 
\begin{align}
	\chi_{[1:l]} = \tr_{l+1 \ldots n}[\chi].
\end{align}
We will also use the convention $\chi_{[1:0]}:= 1$ (the trivial state) and write $\chi_{\{k\}}$ for the reduced state on the $k$-th subsystem. Note that for any $k=1,\ldots,n$ we have
\begin{align}
	\mc D_{\rho'}[\chi_{\{k\}}] = \rho'
\end{align}
by construction.

Now let $A$ be a $n$-dimensional system with orthonormal basis $\ket{k}_A$ and $R$ be a $d$-dimensional system, where $d$ denotes the Hilbert-space dimension of $\rho'$.
Then the catalyst $C$ consists of $n-1$ systems $S_2,\ldots,S_n$ (copies of $S=S_1$) together with the systems $A$ and $R$ and its state $\sigma$ is constructed as
\begin{align}
	\sigma = \frac{1}{n} \sum_{k=1}^n \rho^{\otimes k-1}\otimes \chi_{[1:n-k]} \otimes \proj{k}_A \otimes \frac{\id_R}{d}.
\end{align}
Note that the unitary $V$ above may be interpreted as acting on the systems $S_1\cdots S_n$. 
The unitary $U$ that implements the catalytic transition is now a product of three unitaries: $U=W_3 W_2 W_1$. In the following we define and discuss the action of each of the unitaries.
The first unitary is defined as
\begin{align}
W_1 = V\otimes \proj{n}_A \otimes \id_R + \sum_{k=1}^{n-1} \id_{S_1\cdots S_n} \otimes \proj{k}_A \otimes\id_R.
\end{align}
It has the effect of applying $V$ to $S_1\cdots S_n$ if system $A$ is in state $\ket{n}$ and doing nothing otherwise.
Thus
\begin{align}
	W_1 \rho\otimes\sigma W_1^\dagger &= \frac{1}{n} \left[\chi \otimes \proj{n}_A + \sum_{k=1}^{n-1} \rho^{\otimes k}\otimes \chi_{[1:n-k]}\otimes\proj{k}_A\right]\otimes\frac{\id_R}{d}. 
\end{align}
The unitary $W_2$ first cyclically permutes $S_i$ with $S_{i+1}$ (i.e., $S_{n+1}=S_1$) and then cyclically maps $\ket{k}_A$ to $\ket{k+1}_A$ with $\ket{0}_A=\ket{n}_A$.
If $\tau$ denotes the operators cyclically shfting the subsystems $S_i$, we thus get
\begin{align}
	W_2 W_1 \rho\otimes \sigma W_1^\dagger W_2^\dagger &= \frac{1}{n} \left[\tau\left[\chi \right]\otimes \proj{1}_A + \sum_{k=1}^{n-1} \tau\left[\rho^{\otimes k}\otimes \chi_{[1:n-k]}\right]\otimes\proj{k+1}_A\right]\otimes\frac{\id_R}{d}.
\end{align}
Finally, $W_3 = \frac{1}{d} \sum_{i=1}^d U_i \otimes \id_{S_2\cdots S_n}\otimes\id_A\otimes  \proj{i}_R$, where $\ket{i}_R$ is an orthonormal basis on $R$ and the $U_i$ are chosen such that $\mc D_{\rho'}[X] = \frac{1}{d} \sum_i U_i X U_i^\dagger$. It is always possible to find such unitaries. 
Note that this last unitary only involves systems $S_1$ and $R$. Its effect is to apply the decoherence channel $\mc D_{\rho'}$ to $S$. Importantly, it was shown in Ref.~\cite{Wilming2021} that it leaves the catalyst completely unchanged, since no correlations are build up between $S_2\cdots S_n A$ and $R$ and the state on $R$ remains maximally mixed. This is easy to see by direct computation, which we omit here.  
We thus find that the final state on $S$ is given by:
\begin{align}
	\tr_C[U\rho\otimes\sigma U^\dagger]=\tr_C[ W_3 W_2 W_1 \rho\otimes \sigma W_1^\dagger W_2^\dagger W_3^\dagger ]&= \mc D_{\rho'}[\tr_C[W_2 W_1 \rho\otimes \sigma W_1^\dagger W_2^\dagger]]
\end{align}
and the final state on $C$ is given by
\begin{align}
	\tr_S[U\rho\otimes \sigma U^\dagger]=\tr_S[ W_3 W_2 W_1 \rho\otimes \sigma W_1^\dagger W_2^\dagger W_3^\dagger ]&= 	\tr_S[W_2 W_1 \rho\otimes \sigma W_1^\dagger W_2^\dagger].
\end{align}
To compute the final states, we make use of the identities $\tr_{S_2\cdots S_n}[\tau[\rho^{\otimes k}\otimes \chi_{[1:n-k]}]] = \chi_{\{n-k\}}$ and $\tr_S[\tau[\rho^{\otimes k}\otimes \chi_{[1:n-k]}]]=\rho^{k} \otimes \chi_{[1:n-k-1]}$.
Then
\begin{align}
	\tr_C[U\rho\otimes \sigma U^\dagger] &= \frac{1}{n} \mc D_{\rho'}\left[\chi_{\{n\}} + \sum_{k=1}^{n-1} \chi_{\{n-k\}}\right] = \frac{1}{n} \sum_{k=1}^n \mc D_{\rho'}[\chi_{\{k\}}] = \rho',
\end{align}
where we used that $\mc D_{\rho'}[\chi_{\{k\}}]=\rho'$.
For the final state on $C$ we obtain
\begin{align}
	\tr_S[U\rho\otimes \sigma U^\dagger] &= \frac{1}{n} \left[\chi_{[1:n-1]}\otimes \proj{1}_A + \sum_{k=1}^{n-1} \rho^{\otimes k} \otimes \chi_{[1:n-(k+1)]} \proj{k+1}_A\right]\otimes\frac{\id_R}{d}\\
	&= \frac{1}{n} \sum_{k=1}^n \rho^{\otimes k-1} \otimes \chi_{[1:n-k]} \otimes \proj{k}_A \otimes\frac{\id_R}{d} = \sigma.
\end{align}
An analogous construction and calculation can be done in the classical case, see the Supplemental Material of Ref.~\cite{Wilming2021}.
%\clearpage
\end{document}